\documentclass[preprint2,twocolappendix,x11names]{aastex631}

\shorttitle{Long-lived Protoplanetary Disks}
\shortauthors{Wang et al.}
\usepackage{amsmath}
\usepackage{longtable}
\usepackage{booktabs}
\usepackage{CJKfntef}

\begin{document}

\title{LAMOST Reveals Long-lived Protoplanetary Disks}

\correspondingauthor{Min Fang}
\email{xlwang@hebtu.edu.cn, mfang@pmo.ac.cn}
\correspondingauthor{Wen-Yuan Cui}
\email{cuiwenyuan@hebtu.edu.cn}

\author[0000-0003-2536-3142]{Xiao-Long Wang}
\altaffiliation{Physics Postdoctoral Research Station at Hebei Normal University.}
\affiliation{Department of Physics, Hebei Normal University, Shijiazhuang 050024, People's Republic of China; xlwang@hebtu.edu.cn}
\affiliation{Guo Shoujing Institute for Astronomy, Hebei Normal University, Shijiazhuang 050024, People's Republic of China}
\affiliation{Hebei Advanced Thin Films Laboratory, Shijiazhuang 050024, People's Republic of China}

\author[0000-0001-8060-1321]{Min Fang}
\affiliation{Purple Mountain Observatory, Chinese Academy of Sciences, Nanjing 210023, People's Republic of China}
\affiliation{School of Astronomy and Space Science, University of Science and Technology of China, Hefei 230026, People's Republic of China}

\author[0000-0002-7616-666X]{Yao Liu}
\affiliation{Purple Mountain Observatory, Chinese Academy of Sciences, Nanjing 210023, People's Republic of China}
\affiliation{School of Physical Science and Technology, Southwest Jiaotong University, Chengdu 610031, People's Republic of China}

\author[0000-0002-6388-649X]{Miao-Miao Zhang}
\affiliation{Purple Mountain Observatory, Chinese Academy of Sciences, Nanjing 210023, People's Republic of China}

\author[0000-0003-1359-9908]{Wen-Yuan Cui}
\affiliation{Department of Physics, Hebei Normal University, Shijiazhuang 050024, People's Republic of China; xlwang@hebtu.edu.cn}
\affiliation{Guo Shoujing Institute for Astronomy, Hebei Normal University, Shijiazhuang 050024, People's Republic of China}

\begin{abstract}

While both observations and theories demonstrate that protoplanetary disks are not expected to live much longer than $\sim$10\;Myr, several examples of prolonged disks have been observed in the past. In this work, we perform a systematic search for aged YSOs still surrounded by protoplanetary disks in the M star catalog from the LAMOST archive. We identify 14 sources older than 10\;Myr, still surrounded by protoplanetary disks and with ongoing accretion activities, significantly improving the census of the category known as the Peter Pan disks. The stellar parameters, variability and accretion properties of these objects, as well as their spatial distribution, are investigated. Nearly all of these objects are distributed far away from nearby associations and star forming regions, but show evidence of being members of open clusters. Investigating the correlation between mass accretion rates and stellar masses, we find these long-lived disks accrete at systematically lower levels, compared to their younger counterparts with similar stellar masses. Studying the evolution of mass accretion rates with stellar ages, we find these aged disks follow similar trend as young ones.

\end{abstract}


\section{Introduction} \label{sec:intro}

Circumstellar disks are an inevitable consequence of angular momentum conservation during forming stars through gravitationally collapsing, and it is through the disks that angular momentum transports outward~\citep{Bodenheimer1995ARA&A..33..199B,Williams2011ARA&A..49...67W}. Disks have been observed around young stars of all masses \citep{Williams2011ARA&A..49...67W} and can exhibit a variety of geometry~\citep[including rings and gaps, spirals, and crescents,][]{Bae2023ASPC..534..423B}. Simulations of collapsing molecular cores have shown that disks form rapidly~\citep{Yorke1993ApJ...411..274Y,Hueso2005A&A...442..703H}, and observations indicate a firm upper limit of $\sim$10\;Myr for the longevity of primordial disks surrounding solar-type stars~\citep{Williams2011ARA&A..49...67W}. Observations of the fraction of stars harboring protoplanetary disks in clusters of different ages have shown that circumstellar disks dissipate within a few million years, with a typical lifetime of $\sim$2$-$5\;Myr and a maximum lifetime of 10$-$20\;Myr~\citep{Haisch2001ApJ...553L.153H,Mamajek2009AIPC.1158....3M,Ribas2014A&A...561A..54R,Pecaut2016MNRAS.461..794P}. \citet{Ribas2014A&A...561A..54R} found characteristic timescales of 4$-$6\;Myr and 2$-$3\;Myr for primordial disks probed at 22$-$24\;$\rm\mu m$ and 3.4$-$12\;$\rm\mu m$, respectively, and there is a trend that circumstellar disks probed at longer wavelength live longer, in agreement with the inside-out disk clearing scenario~\citep{Shu1993Icar..106...92S}. In addition, the mass of the hosting star plays a major role in driving disk formation, evolution, and dissipation. Early surveys of nearby star-forming regions by the Spitzer telescope have suggested that protoplanetary disks surrounding low-mass stars persist longer than around solar- and high-mass stars~\citep{Carpenter2006ApJ...651L..49C,Lada2006AJ....131.1574L,Hernandez2007ApJ...662.1067H,Hernandez2007ApJ...671.1784H,Kennedy2009ApJ...695.1210K}. \citet{Mamajek2009AIPC.1158....3M} found that the disk fraction decay timescale varies with the mass of the hosting star. \citet{Luhman2012ApJ...758...31L} found $\sim$25\% of M5-L0 members in the Upper Sco region possess inner disks, while this fraction decreases to only 10\% for B-G stars. And~\citet{Fang2012A&A...539A.119F} found substantially lower inner disk frequencies in clusters harboring extremely massive stars. \citet{Ribas2015A&A...576A..52R} showed that disks surrounding high-mass stars dissipate up to twice as fast as surrounding low-mass ones.

In light of the above observational results, several examples of protoplanetary disks accreting at ages greater than $\sim$10\;Myr have been identified~\citep{Mamajek2002AJ....124.1670M,Moor2011ApJ...740L...7M,Zuckerman2012ApJ...758...77Z,Silverberg2016ApJ...830L..28S,Murphy2018MNRAS.476.3290M,Silverberg2020ApJ...890..106S,Lee2020MNRAS.494...62L}. \citet{Lee2020MNRAS.494...62L} tabulated a sample of 15 such cases, with ages ranging from $\sim$10\;Myr to $\sim$70\;Myr, and associated these anomalies to nearby associations. These unusual anomalies have been termed the ``Peter Pan'' disks~\citep{Silverberg2020ApJ...890..106S}. Besides these nearby Peter Pan disks, other long-lived accretion disks have also been observed outside the immediate solar neighbourhood~\citep{Currie2007ApJ...659..599C,Currie2009AJ....138..703C,Beccari2010ApJ...720.1108B,Spezzi2012MNRAS.421...78S,De-Marchi2013ApJ...775...68D,De-Marchi2013MNRAS.435.3058D}. In spite of the small number of Peter Pan disks, these ``extremely old'' PMS stars harboring primordial disks can greatly improve our understanding to the theory of planet formation~\citep{Greaves2010MNRAS.407.1981G,Najita2014MNRAS.445.3315N,Manara2018A&A...618L...3M,Pfalzner2019ApJ...874L..34P} and provide upper limit on the lifetimes of gaseous disks. The existence of these old, low-mass accreting stars indicates that at least some low-mass stars can retain their gas reservoirs much longer than previously recognized.

Follow up observations and simulations have been performed to explore the properties and origin of these rare, long-lived primordial disks. \citet{Laos2022ApJ...935..111L} observed six Peter Pan disks using Chandra, and suggested that Peter Pan disks may be a consequence of the low-level far-UV radiation incident on the disks surrounding low-mass stars. By exploring how different mass loss processes limit the maximum lifetimes of protoplanetary disks, \citet{Wilhelm2022MNRAS.509...44W} demonstrated that Peter Pan disks can only occur around M dwarfs.

Studying a large sample of isochronally aged YSOs still surrounded by primordial disks is of vital importance for understanding the physical origin of these anomalies and constraining the timescale for planet formation. In addition, these extremely old cases provide good benchmarks through which to study late-stage evolution of protoplanetary disks. In this study, we perform a search for extremely aged YSOs still surrounded by primordial disks, based mainly on LAMOST spectroscopic data and WISE photometry. We describe the data sets in Section~\ref{sec:data}, and the selection of our sample is described in Section~\ref{sec:select}. The source properties are determined in Section~\ref{sec:target_prop}. We present our discussion in Section~\ref{sec:discussion} and summary in Section~\ref{sec:summary}.

%%%%%%%%%%%%%%%%%%%%%%%%%%%%%%%%%%%%%%%%%%%%%%%%%%%%%%%%%%%%%%%%%%%%%%%%%%%%%%%%%%%%%%%%%%%%

\section{Data Sets}\label{sec:data}

The main data sets used in this work include the spectroscopic data from the LAMOST\footnote{The Large Sky Area Multi-Object Fiber Spectroscopic Telescope, also called the Guoshoujing Telescope.} survey~\citep{Cui2012RAA....12.1197C}, the astrometric measurements from the Gaia satellite~\citep{Gaia-Collaboration2016A&A...595A...1G}, and multiband photometry from the Pan-STARRS1 survey~\citep[PS1,][]{Hodapp2004AN....325..636H}, the Two Micron All Sky Survey~\citep[2MASS,][]{Skrutskie2006AJ....131.1163S} and the Wide-field Infrared Survey Explorer~\citep[WISE,][]{Wright2010AJ....140.1868W}.

There are $\sim$0.88 million spectra of M type stars, corresponding to more than 0.66 million unique sources, in the 10th release of the LAMOST survey (LAMOST DR10\footnote{\url{http://www.lamost.org/dr10/}}). The M star catalog is firstly cross-matched with Gaia DR3~\citep{Gaia-Collaboration2023A&A...674A...1G}. The matched spectra are further matched to the PS1 DR1~\citep{Chambers2016arXiv161205560C}, the 2MASS All-Sky Point Source Catalog~\citep{Skrutskie_2MASS_IPAC}, and the AllWISE catalog~\citep{Wright_AllWISE_IPAC}. The full SED (from optical to infrared) is constructed for each star, and only stars within 1\;kpc (i.e., $\varpi>1\;\rm mas$) are retained for further analysis. During cross-matching, photometry with magnitude errors $\ge$0.2\;mag are omitted. For the Gaia catalog, we have also omitted sources with parallax/parallax\_error$\le$5, and only sources with $ruwe<1.4$ \citep{Gaia-Collaboration2021A&A...649A...1G,Lindegren2018RUWE} are retained for subsequent analysis. For the PS1 catalog, saturated photometry \citep[$<$13.5\;mag for $gri$-bands, $<$13.0\;mag for $z$-band, and $<$12.0\;mag for $y$-band,][]{Magnier2013ApJS..205...20M} are omitted as well. To better constrain the stellar properties, only sources with at least two valid PS1 bands (i.e., PS1 bands that are not removed due to large magnitude errors or due to saturation) are retained. The preprocessed catalog is designated the ``full catalog''.

%%%%%%%%%%%%%%%%%%%%%%%%%%%%%%%%%%%%%%%%%%%%%%%%%%%%%%%%%%%%%%%%%%%%%%%%%%%%%%%%%%%%%%%%%%%%

\section{Target Selection}\label{sec:select}

\subsection{Identification of Disked Objects}\label{sec:select_disk}

\begin{figure*}[!t]
\includegraphics[width=\textwidth]{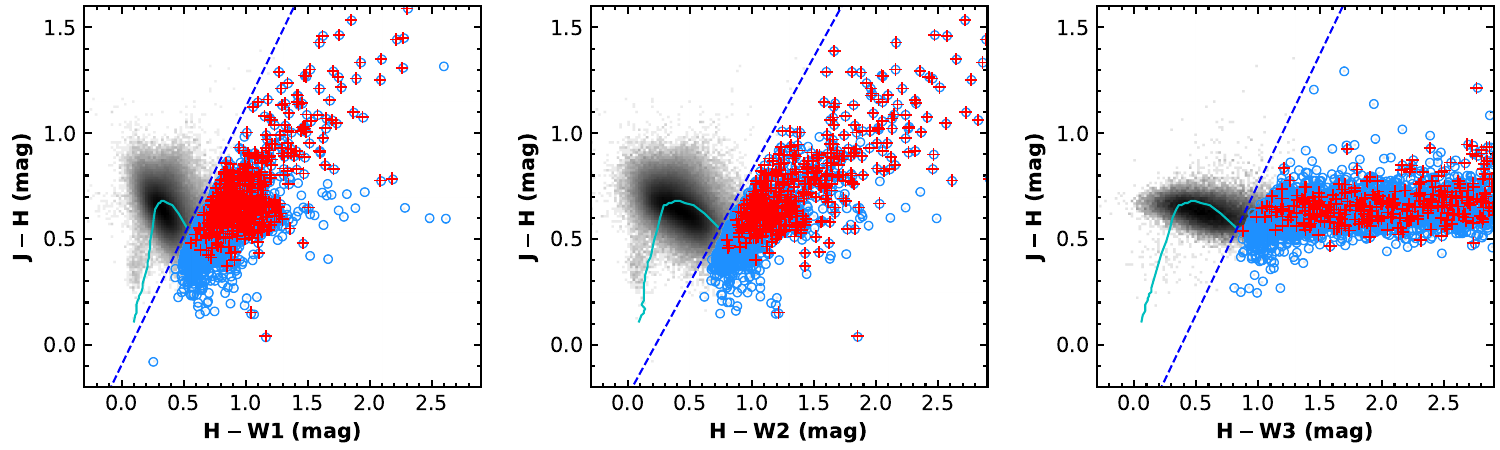}
\caption{Infrared color-color plots used to select candidates displaying infrared excess emission. In each panel, the solid cyan curve denotes the intrinsic colors of 5$-$30\;Myr old stars \citep{Pecaut2013ApJS..208....9P}. The dashed blue line displays the extinction direction \citep{Wang2019ApJ...877..116W}, and is used as the border line separating disked objects from diskless ones. The background gray image displays the distribution of the full catalog. The blue circles are candidates showing excess emission in the color-color plots, and the red pluses mark our working sample comprising of 855 genuine disked objects (see Section~\ref{sec:select_disk} for selecting genuine disked objects).}\label{fig:CCD}
\end{figure*}

The main goal of this work is to search for aged YSOs (older than $\sim$10\;Myr) still surrounded by primordial disks in the solar vicinity (within 1\;kpc of the Sun). For this purpose, we firstly identify sources displaying infrared excess emission using three color-color plots (Figure~\ref{fig:CCD}). By comparing observed infrared colors with the intrinsic colors of 5$-$30\;Myr old stars \citep{Pecaut2013ApJS..208....9P}, we select sources obey any of the following constraints as candidate disked objects: (1) $[J-H]+\sigma_{[J-H]}< 1.22\times([H-W1]-\sigma_{[H-W1]}-0.53)+0.55$; (2) $[J-H]+\sigma_{[J-H]}< 1.07\times([H-W2]-\sigma_{[H-W2]}-0.74)+0.55$; (3) $[J-H]+\sigma_{[J-H]}< 1.23\times([H-W3]-\sigma_{[H-W3]}-0.83)+0.55$. With these criteria, we select 7201 sources as disked candidates from the full catalog.

Since the disked candidates are identified based on WISE photometry, the reliability of the WISE photometry affects our results significantly. Although we have applied quality cuts on the WISE photometry (see Section~\ref{sec:data}), there still are sources with unreliable photometry due to fake source detection or due to contamination from nearby or background sources \citep{Koenig2014ApJ...791..131K}. For this reason, the WISE images of the candidates are inspected visually one-by-one, and unreliable photometry are removed. Necessity of checking the WISE images are demonstrated in Appendix~\ref{app:wise_image}.

In addition, while the initial selection of disked candidates is characterized by their red infrared colors, we can not rule out the possibility of the sources being reddened diskless stars. For this reason, the candidates are firstly spectral typed using LAMOST spectra (see Section~\ref{sec:spt}), and SED fitting (see Section~\ref{sec:sedfitting}) is performed to further assess the existence of infrared excess. Following \citet{Luhman2022AJ....163...25L}, if a star displays excess emission in a given band, but a reliable detection in any band at a longer wavelength is consistent with the best-fitted stellar photosphere, the star is removed from the sample of disked candidates. Finally, we identify 855 sources showing significant infrared excess emission $5\sigma$ above the stellar photosphere in WISE bands, indicating the presence of circumstellar disks. This sample of 855 disked objects constitutes the ``working sample'' of this study, and they are highlighted as red pluses in Figure~\ref{fig:CCD}. In the next subsection, we will identify isochronally old disks from the working sample, and the disk properties are discussed in Section~\ref{sec:disktype}.

\subsection{Identification of Aged Disks}\label{sec:select_old}

\setlength{\tabcolsep}{4pt}
\begin{longrotatetable}
\centering
\movetabledown=10ex
\begin{deluxetable*}{lccccccccccccccc}
\centering
\tablecaption{Properties of the Newly discovered Peter Pan disks\label{tab:PeterPanDisk}}
% \tablenum{1}
\tablewidth{0pt}
\tabletypesize{\scriptsize}
\tablehead{
\colhead{GaiaDR3}     & \colhead{RA}          & \colhead{DEC}                & \colhead{SPT}           & \colhead{$T_{\rm eff}$} & \colhead{$\log L_{\rm bol}$} & %
\colhead{$M_{\star}$} & \colhead{Age}         & \colhead{$\rm EW_{H\alpha}$} & \colhead{$\log\dot{M}$} & \colhead{variable}      & \colhead{DiskType}           & %
\colhead{Cluster}     & \colhead{ClusterType} & \colhead{Cloud}              & \colhead{CloudType}\\
\colhead{}      & \colhead{deg}                        & \colhead{deg} & \colhead{}    & \colhead{(K)} & \colhead{($L_{\odot}$)} & \colhead{($M_{\odot}$)} & \colhead{(Myr)} & %
\colhead{(\AA)} & \colhead{($M_{\odot}\;\rm yr^{-1}$)} & \colhead{}    & \colhead{}    & \colhead{}    & \colhead{}              & \colhead{}              & \colhead{}
}
\startdata
155649614856576            &  45.784809   &  0.911900   &  $\rm M4.2\pm0.7$  &  3356  &  -2.28  &  0.41  &  $>$50$^{b}$  &  16.9   &  -12.44  &  Y  &  FULL     &  $\cdots$  &  $\cdots$  &  $\cdots$  &  $\cdots$ \\ 
163182888662060928         &  63.549555   &  28.198063  &  $\rm M7.2\pm1.6$  &  3212  &  -1.32  &  0.40  &  21.6         &  347.1  &  -10.17  &  Y  &  FULL     &  H2861     &  6         &  Taurus    &  3        \\ 
169126642366111744         &  62.307499   &  31.609006  &  $\rm M4.0\pm0.9$  &  3660  &  -1.00  &  0.66  &  32.9         &  187.0  &  -9.22   &  Y  &  FULL     &  $\cdots$  &  $\cdots$  &  $\cdots$  &  $\cdots$ \\ 
462182014838117120         &  51.426696   &  58.958308  &  $\rm M1.0\pm1.5$  &  4080  &  -0.74  &  0.73  &  17.2         &  53.8   &  -8.82   &  N  &  FULL     &  $\cdots$  &  $\cdots$  &  $\cdots$  &  $\cdots$ \\ 
2162887638405193216        &  313.541882  &  44.117787  &  $\rm K9.4\pm1.8$  &  3666  &  -1.37  &  0.53  &  $>$50$^{b}$  &  86.3   &  -9.49   &  N  &  FULL     &  $\cdots$  &  $\cdots$  &  $\cdots$  &  $\cdots$ \\ 
3209547627721019904        &  83.789441   &  -4.953254  &  $\rm M4.6\pm0.6$  &  3257  &  -1.20  &  0.45  &  17.7         &  192.9  &  -9.15   &  Y  &  FULL     &  H4503     &  2         &  $\cdots$  &  $\cdots$ \\ 
3213833322184248320        &  78.132566   &  -3.016444  &  $\rm M3.0\pm0.3$  &  3797  &  -0.78  &  0.71  &  14.2         &  95.4   &  -9.06   &  Y  &  TD       &  H5101     &  3         &  $\cdots$  &  $\cdots$ \\ 
3214473272311601152        &  77.439366   &  -3.155679  &  $\rm M3.2\pm0.7$  &  3795  &  -0.74  &  0.71  &  11.6         &  44.8   &  -8.98   &  Y  &  FULL     &  H5101     &  5         &  $\cdots$  &  $\cdots$ \\ 
3223542525253775104        &  82.551684   &  1.805948   &  $\rm M3.8\pm0.9$  &  3452  &  -1.13  &  0.56  &  25.1         &  174.2  &  -9.04   &  Y  &  FULL     &  H12       &  6         &  $\cdots$  &  $\cdots$ \\ 
3241216624914091136        &  79.299007   &  7.039780   &  $\rm M3.2\pm0.5$  &  3783  &  -0.88  &  0.71  &  24.7         &  35.9   &  -9.70   &  Y  &  FULL     &  H4978     &  6         &  $\cdots$  &  $\cdots$ \\ 
3308700559817832576$^{a}$  &  72.003657   &  14.665950  &  $\rm M4.8\pm0.4$  &  2750  &  -1.77  &  0.16  &  12.4         &  72.8   &  -10.07  &  Y  &  FULL     &  $\cdots$  &  $\cdots$  &  $\cdots$  &  $\cdots$ \\ 
3319249446173297664        &  86.651606   &  3.100752   &  $\rm M4.4\pm0.5$  &  2686  &  -1.89  &  0.14  &  14.4         &  51.7   &  -10.57  &  Y  &  EVOLVED  &  $\cdots$  &  $\cdots$  &  $\cdots$  &  $\cdots$ \\ 
3319360599927089024        &  88.200068   &  3.748879   &  $\rm M3.6\pm0.2$  &  2897  &  -1.85  &  0.20  &  29.9         &  15.4   &  -11.11  &  N  &  EVOLVED  &  $\cdots$  &  $\cdots$  &  $\cdots$  &  $\cdots$ \\ 
3415666239289777664$^{a}$  &  79.511923   &  23.453350  &  $\rm M5.2\pm0.5$  &  2623  &  -1.97  &  0.12  &  13.6         &  22.2   &  -11.27  &  Y  &  EVOLVED  &  H5496     &  6         &  $\cdots$  &  $\cdots$ \\
\enddata
\tablecomments{\textbf{Note}: Only a portion of columns are displayed here for display purpose. The full table and column descriptions can be found in Table~\ref{tab:AllDiskedObjects}. $^{a}$Sources are also listed as members of the 32~Ori association by \citet{Luhman2022AJ....164..151L}. $^{b}$Sources well below the 50\;Myr isochrone are assigned ages of $>$50\;Myr.}
\end{deluxetable*}
\end{longrotatetable}

In this subsection we identify isochronally old disked objects by comparing source locations in the Hertzsprung-Russell (H-R) diagrams with the 10\;Myr isochrone from the PARSEC stellar model \citep{Bressan2012MNRAS.427..127B}. Since our working sample comprises of bona fide disked objects, disk scattering may play important roles in shaping the emergent radiation. For this reason, we construct two types of H-R diagrams (Section~\ref{sec:sedfitting}), with and without correcting for disk scattering. To be stringent, only sources below the 10\;Myr isochrone in both H-R diagrams are identified as old. Finally, we identify 526 isochronally aged YSOs still surrounded by circumstellar disks. In the following, we will denote these aged disked YSOs as the ``old population'', and the remaining 329 disked objects as the ``young population''. Of the old population, 50 objects display H$\alpha$ in emission. Fourteen of them are characterized as classical T Tauri stars (CTTSs) in Section~\ref{sec:acc}, and the remaining 36 are weak-line T Tauri stars (WTTSs). Following \citet{Silverberg2020ApJ...890..106S}, the 14 isochronally aged CTTSs are termed the Peter Pan disks, and all of them are newly discovered ones. 

Target properties of these disked objects are characterized in Section~\ref{sec:target_prop}. In Table~\ref{tab:PeterPanDisk}, We list some of the derived properties of the newly discovered Peter Pan disks for display purpose. The full information for the whole working sample of disked objects (including the newly discovered Peter Pan disks) are provided in Appendix~\ref{app:AllDiskedObjects}.

%%%%%%%%%%%%%%%%%%%%%%%%%%%%%%%%%%%%%%%%%%%%%%%%%%%%%%%%%%%%%%%%%%%%%%%%%%%%%%%%%%%%%%%%%%%%

\section{Target Properties}\label{sec:target_prop}

\subsection{Spectral Type}\label{sec:spt}

\begin{figure}[!t]
    \centering
    \includegraphics[width=\columnwidth]{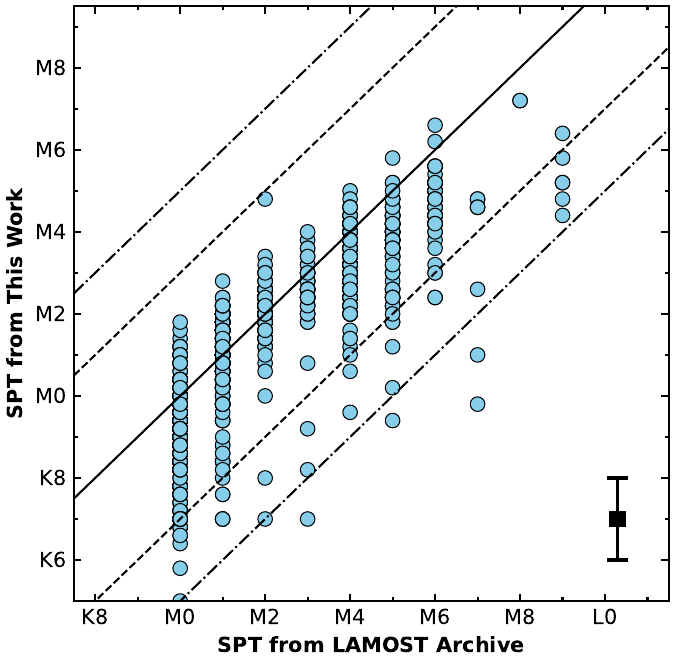}
    \caption{Comparison of the spectral types derived in this work and that from the LAMOST archive. The solid line is the line of equality. The dashed and dash-dotted lines show differences of 3 and 5 subtypes, respectively. In the bottom-right corner, we display the typical uncertainties of our spectral typing ($\sim$1.0 subtypes).}
    \label{fig:SPThere_vs_SPTlamost}
\end{figure}

In this section, we estimate spectral types for the 855 disked objects. Due to the complexity of classifying late-type stars, the LAMOST pipeline~\citep{Wu2011A&A...525A..71W,Luo2015RAA....15.1095L} may give incorrect spectral types. To address this issue, we reclassify the sources using the classification scheme from~\citet{Fang2020ApJ...904..146F}. An excess flux is added to the spectral template to account for the filling effect on the photospheric absorption lines due to excess emission from the accretion shocks when necessary. Following \citet{Herczeg2014ApJ...786...97H}, the accretion continuum is assumed to be constant. We fit both veiled and nonveiled templates to our spectra, and perform visual inspection of the fitting results. The depths of the absorption features of the templates are compared to that of the observed spectra and the final types are determined to be the one better matching the observed depths. In Appendix~\ref{app:spec_example}, we display two examples demonstrating the spectral typing and the reader is referred to~\citet{Fang2020ApJ...904..146F} for more details on the spectral classification. From the comparison, for the source \textit{Gaia DR3 147605282796916992}, the observed depths of the absorption features around $6150\;\rm\AA$ are better matched by the nonveiled fitting (blue), while for \textit{Gaia DR3 121394677937069184}, the veiled fitting (red) matches the observed depths better. In Figure~\ref{fig:SPThere_vs_SPTlamost}, we compare the spectral types derived in this work and that from the LAMOST archive. In most cases, the two are consistent within 3 subtypes (the dashed lines in Figure~\ref{fig:SPThere_vs_SPTlamost}). There are 4 K/M stars differ more than 5 subtypes (dash-dotted lines in Figure~\ref{fig:SPThere_vs_SPTlamost}). Three of them also have literature types from various studies and our types are generally consistent with that from the literature. These cases include, the source \textit{Gaia DR3 147605282796916992} classified as M7 by the LAMOST pipeline, as M1 by us, and as M0.5 by \citet{Liu2021ApJS..254...20L}, the sources \textit{Gaia DR3 216573420958954112} and \textit{Gaia DR3 3219615241517223936} classified as M5 and M3 by the LAMOST pipeline, as K9.4 and K7.0 by us, and as M1.7 and M1.8 by \citet{Birky2020ApJ...892...31B}.

Although we start our search with the LAMOST M star catalog, there are several sources reclassified as K types. This is not surprising because the LAMOST pipeline is best suited to classify main-sequence stars and no veiling effect is considered in the pipeline, but most of the 855 disked objects are strong accretors (Section~\ref{sec:acc}).

%%%%%%%%%%%%%%%%%%%%%%%%%%%%%%%%%%%%%%%%%%%%%%%%%%%%%%%%%%%%%%%%%%%%%%%%%%%%%%%%%%%%%%%%%%%%

\subsection{Effective Temperatures, Bolometric Luminosities, Stellar Ages and Masses}\label{sec:sedfitting}

\begin{figure*}[!t]
    \centering
    \includegraphics[width=\textwidth]{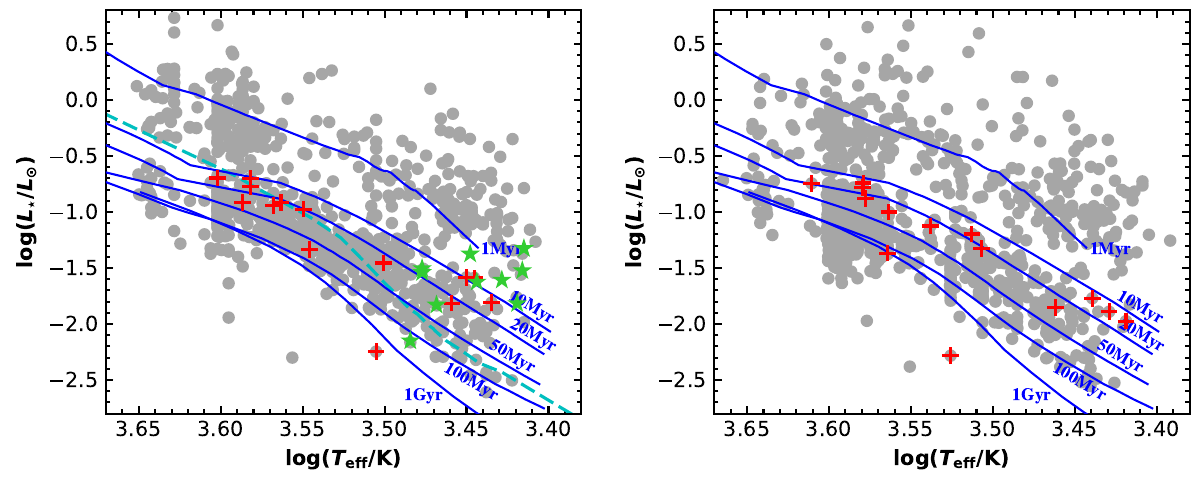}\hfill%
    \caption{H-R diagram of the identified disked objects (gray circles). Left: the stellar luminosities and effective temperatures are determined by fitting stellar photospheres to the observed optical and NIR photometry. The blue solid curves are isochrones from the PARSEC stellar model~\citep{Bressan2012MNRAS.427..127B} with corresponding ages labeled. The 14 newly discovered Peter Pan disks are marked with additional red pluses. Right: similar as the left panel, but the stellar luminosities and effective temperatures are obtained by fitting disked models to the full SEDs. Also displayed in the left panel are the Peter Pan disks (green stars) from \citet{Lee2020MNRAS.494...62L}, and the 10\;Myr isochrone (cyan dashed curve) from \citet{Baraffe2015A&A...577A..42B}. }
    \label{fig:HRD}
\end{figure*}

In this section, we estimate the effective temperatures and bolometric luminosities for the 855 disked objects through SED fitting, to construct the H-R diagrams (Figure~\ref{fig:HRD}). We fit two sets of model SEDs to the observed photometry. The first set of model SEDs include radiation from pure stellar photospheres only, and in the second set, both the central stellar sources and circumstellar disks contribute to the emergent radiation. More details of constructing the model SEDs are described in \citet{Robitaille2017A&A...600A..11R}. The first set of model SEDs are fitted to only $grizy$ and $J$-bands photometry of the targets, while the second set are fitted to the full SEDs.

The model SEDs are convolved with common filters including $grizy$ bands of the PS1 survey, the $JHK$ bands of the 2MASS survey, and the four infrared bands of the WISE survey. The convolved fluxes are then converted to magnitudes, and we perform chi-square fitting in magnitude space. Since we have determined spectral types for our targets in Section~\ref{sec:spt}, only a subset SEDs with temperatures within 500\;K of that corresponding to the spectral types are fitted to the targets. The scaling relation from~\citet{Fang2017AJ....153..188F} is adopted to convert spectral types to effective temperatures. The reason that we don't fix the effective temperatures to the values corresponding to the spectral types is to account for possible uncertainties in the spectral typing and in the scaling relation.

For a given source and for a specified model SED in the subset, the $\chi^{2}$ function to be optimized is
\begin{equation}\label{eq:chi2}
\chi^{2}=\dfrac{1}{N}\sum_{i=1}^{N}\left[
\dfrac{m_{i}-\left(M_{i}+C+A_{V}F^{\rm ext}_{i}\right)}{\sigma_{i}}
\right]^{2},
\end{equation}
where $N$ denotes the number of filters used in the fitting, $m_{i}$ is the observed magnitude, $\sigma_{i}$ is the corresponding magnitude error, $M_{i}$ is the magnitude corresponding to the model flux at the stellar surface and $F^{\rm ext}_{i}$ is the extinction coefficient at the corresponding filter. The fitted parameters are $A_{V}$ and $C$, where $A_{V}$ is the extinction toward the source, and $C$ is a scaling factor, related to the angular size of the source as $C=-5\log(R_{\star}/d)$. In this work, we adopt the extinction law from~\citet{Cardelli1989ApJ...345..245C}. Minimizing the $\chi^{2}$ function and running through the subset, we obtain the best fitted $A_{V}$, $C$, and effective temperature ($T_{\rm eff}$). The stellar radius ($R_{\star}$) is determined with the best-fitted scaling factor $C$ and the bolometric luminosity is calculated using the Stefan-Boltzmann law:
\begin{equation}
L_{\rm bol}=4\pi R_{\star}^{2}\sigma T_{\rm eff}^{4},
\end{equation}
where $\sigma$ is the Stefan-Boltzmann constant.

\begin{figure*}[!t]
    \centering
    \includegraphics[width=\textwidth]{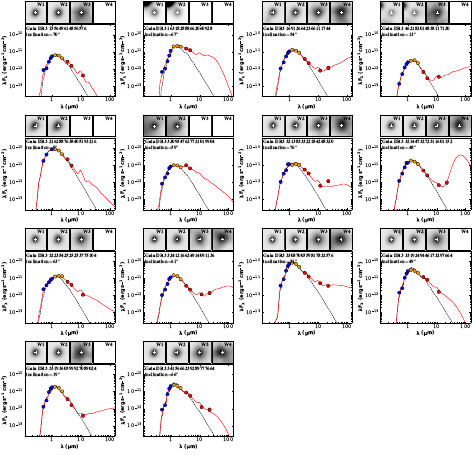}
    \caption{SEDs and WISE images of the 14 newly discovered Peter Pan disks. For each source, the main panel shows the SED of the target, with blue, orange and red circles represent optical, NIR and WISE photometry, respectively. The black dashed line represents the best-fitted stellar photosphere, and the red solid line is the best-fitted disked model. The source names and the best-fitted inclination angles are labeled. The upper panels show the WISE images of the source, and only bands with reliable photometry are displayed.}
    \label{fig:sed_old_ctts}
\end{figure*}

We display the observed SEDs and the corresponding best-fitted SEDs for the 14 Peter Pan disks in Figure~\ref{fig:sed_old_ctts}. As shown in the figure, all but two of the newly discovered Peter Pan disks have best-fitted inclination angles less than 70$^{\circ}$, hinting that most of them are likely to be old disks, instead of being young disks viewed edge-on. Li\,{\footnotesize I}\,$\lambda6707\,\rm\AA$ absorption line is a good indicator of stellar youth. We inspected the LAMOST spectra, but the spectra are so noisy that no clear Li\,{\footnotesize I}\,$\lambda6707\,\rm\AA$ absorption lines are detected. While the resolving power of LAMOST ($\sim$1800) is incapable of separating the stellar emission from the telluric lines, future high resolution spectroscopy of the [O\,{\footnotesize I}] $\lambda6300\,\rm\AA$ forbidden lines may help to constrain their disk inclination \citep[e.g.,][]{Simon2016ApJ...831..169S,Fang2018ApJ...868...28F,Banzatti2019ApJ...870...76B}.

With derived effective temperatures and bolometric luminosities, we construct the H-R diagrams (Figure~\ref{fig:HRD}), and stellar masses ($M_{\star}$) and ages are determined by comparing source locations in the H-R diagram with the PARSEC stellar model~\citep{Bressan2012MNRAS.427..127B}. Although we determine two sets of effective temperatures and bolometric luminosities, and two H-R diagrams are constructed, we determine stellar masses and ages using the one corrected for disk scattering, since disks contribute significantly to the emergent radiation of these disked objects. The masses and ages are provided in Table~\ref{tab:AllDiskedObjects} for the whole working sample.

%%%%%%%%%%%%%%%%%%%%%%%%%%%%%%%%%%%%%%%%%%%%%%%%%%%%%%%%%%%%%%%%%%%%%%%%%%%%%%%%%%%%%%%%%%%%

\subsection{Variability Properties}\label{sec:ztf_var}

\begin{figure}[!t]
    \centering
    \includegraphics[width=\columnwidth]{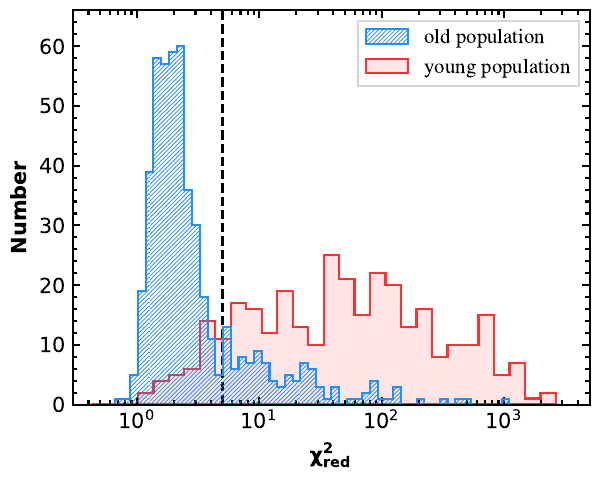}
    \caption{Histogram shows the distribution of the reduced $\chi^{2}$ of the lightcurves. The old and young populations are shown as blue hatched and red filled histograms, respectively. The black vertical dashed line marks the threshold distinguishing between variables and non-variables (i.e., $\chi_{\rm red}^{2}=5$).}
    \label{fig:ztf_var}
\end{figure}

We collect $r$-band time series photometry from the ZTF survey~\citep{Kulkarni2018ATel11266....1K}\footnote{By the time we start this work, the latest release is the ZTF DR21.} for the 855 disked objects. For each lightcurve, observations with \texttt{catflags=32768} are ignored to avoid contamination from clouds or the moon. Outlier measurements $5\sigma$ away from the median magnitude are also removed. Only lightcurves with more than 10 valid measurements are retained for our analysis. Finally, we retrieve 822 lightcurves for the disked objects. Variability properties are determined for the disked objects. We use the reduced $\chi^{2}$ metric~\citep{Sokolovsky2017MNRAS.464..274S}
\begin{equation}
\chi_{\rm red}^{2}=\dfrac{1}{N-1}\sum_{i=1}^{N}\dfrac{(m_{i}-\overline{m})^{2}}{\sigma_{i}^{2}},
\end{equation}
to assess the variability property, where $N$ is the number of measurements, $m_{i}$ is a magnitude measurement and $\sigma_{i}$ is the corresponding measurement uncertainty. $\overline{m}$ is the weighted mean magnitude defined as
\begin{equation}
\overline{m}=\left.\sum_{i=1}^{N}\dfrac{m_{i}}{\sigma_{i}^{2}}\middle/\sum_{i=1}^{N}\dfrac{1}{\sigma_{i}^{2}}\right.,
\end{equation}
The distribution of $\chi_{\rm red}^{2}$ is displayed in Figure~\ref{fig:ztf_var}. Similar as in~\citet{Rebull2014AJ....148...92R} and to be conservative, we identify these having $\chi_{\rm red}^{2}>5$ as variables. Nearly all (286/319=90\%) stars of the young population are variables, while only 21\% (105/503) of the old population are variables. All but 3 of the 14 newly discovered Peter Pan disks are also variables. The variability fraction of the young disked population is similar to that found for disked YSOs in the Perseus cloud \citep{Wang2023RAA....23g5015W}.

%%%%%%%%%%%%%%%%%%%%%%%%%%%%%%%%%%%%%%%%%%%%%%%%%%%%%%%%%%%%%%%%%%%%%%%%%%%%%%%%%%%%%%%%%%%%

\subsection{Accretion Properties}\label{sec:acc}

\begin{figure}[!t]
    \centering
    \includegraphics[width=\columnwidth]{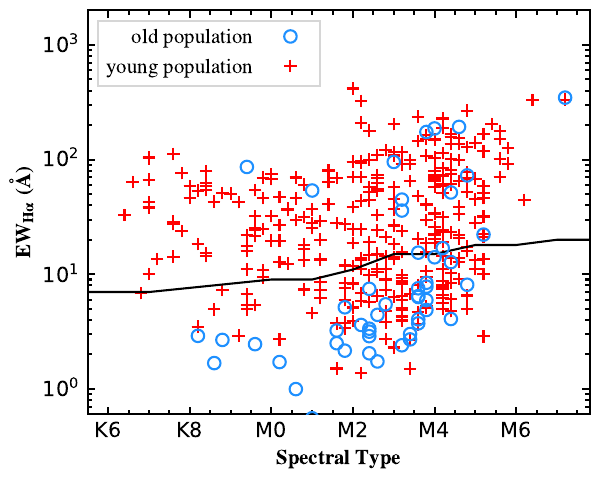}
    \caption{Equivalent width of H$\alpha$ emission line as a function of spectral type. The young population is displayed as red pluses and the old population as blue circles. The solid curve is the dividing line separating strong accretors from weak ones ~\citep{Fang2009A&A...504..461F}.}
    \label{fig:SPT_vs_EW}
\end{figure}

\begin{figure*}[!t]
    \centering
    \includegraphics[width=\textwidth]{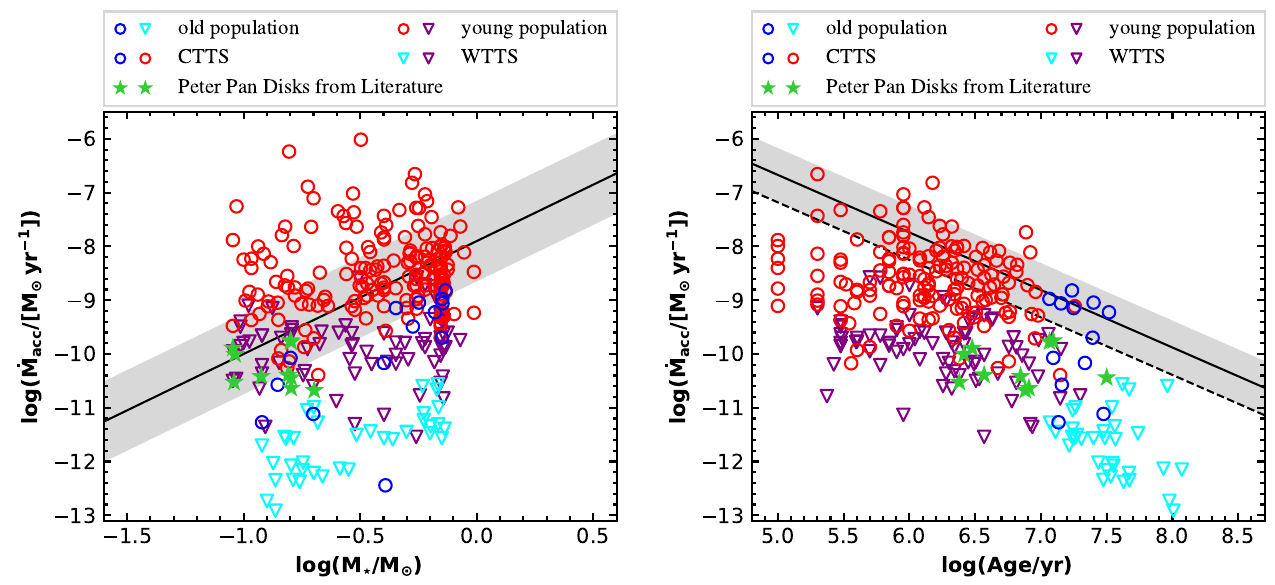}
    \caption{Accretion rates versus stellar masses (left) and ages (right). In each panel, Young and old accretors are shown as red and blue circles, respectively. Non-accreting stars are shown as downward triangles, with purple for young population and cyan for old population. The green star symbols are Peter Pan disks tabulated in~\citet{Lee2020MNRAS.494...62L}. In the left panel, the solid line and gray shaded area show the best linear fit and the corresponding 1-$\sigma$ scatter (0.75\;dex) from~\citet{Hartmann2016ARA&A..54..135H}. In the right panel, the solid line and gray shaded area show the best-fitted relation for stars with $M_{\star}=0.7\;M_{\odot}$ and the corresponding 1-$\sigma$ scatter (0.5\;dex). The dashed line is the same relation corrected for stellar mass ($M_{\star}=0.4\;M_{\odot}$, the typical stellar mass of our sample) using the relation in the left panel. To perform the correction, the solid line is downward by $\sim$0.5\;dex according to the difference of mass accretion rates of 0.4 and $0.7\;M_{\odot}$ stars using the relation in the left panel.}
    \label{fig:Macc_vs_star}
\end{figure*}

Accreting YSOs are generally characterized by strong and broad emission lines in their optical to near-infrared spectra~\citep{Hartmann1994ApJ...426..669H,Muzerolle1998AJ....116.2965M,Muzerolle1998AJ....116..455M}. In this section, we use the H$\alpha$ emission lines to study the accretion activities of the sources. We use the equivalent width of H$\alpha$ emission lines ($\rm EW_{H\alpha}$) to separate CTTSs from WTTSs (Figure~\ref{fig:SPT_vs_EW}). Following the prescription in~\citet{Fang2009A&A...504..461F}, we classify as strong accretors these sources with $\rm EW_{H\alpha}\ge3\;\AA$ for K0-K3 stars, $\rm EW_{H\alpha}\ge5\;\AA$ for K4 stars, $\rm EW_{H\alpha}\ge7\;\AA$ for K5-K7 stars, $\rm EW_{H\alpha}\ge9\;\AA$ for M0-M1 stars, $\rm EW_{H\alpha}\ge11\;\AA$ for M2 stars, $\rm EW_{H\alpha}\ge15\;\AA$ for M3-M4 stars, $\rm EW_{H\alpha}\ge18\;\AA$ for M5-M6 stars, $\rm EW_{H\alpha}\ge20\;\AA$ for M7-M8 stars. Among the 855 disked objects,  352 objects display prominent H$\alpha$ emission lines in their LAMOST spectra. Of these H$\alpha$ emitters, 224 objects are characterized as CTTSs and the remaining 128 objects are WTTSs. The source \textit{Gaia DR3 159614423673903616}, with spectral type of M4 and $\rm EW_{H\alpha}=22\;\AA$, is marginally above the thresholds. Considering that its LAMOST spectrum is very noisy, that it only shows very weak excess in $W3$, and its quasi-periodic lightcurve, we reclassify it as a WTTS. The remaining 503 disked objects, of which the spectra are too noisy to detect clear H$\alpha$ emission lines, are assigned WTTSs. Of the 329 young objects, 210 ($\sim$64\% of the young population) are characterized as CTTSs, while only $\sim$3\% (14/526) of the old population are CTTSs. These fractions are consistent with the general trend of decreasing accretion rates with increasing stellar ages \citep[][and references therein]{Hartmann2016ARA&A..54..135H}.

Besides studying the accretion activity qualitatively, we also determine the mass accretion rates for quantitative study. The continuum flux around the H$\alpha$ line is obtained by interpolating the best-fitted model SED, and the line luminosity ($L_{\rm H\alpha}$) is determined by multiplying the continuum flux by the equivalent width. The emission line luminosity is further converted to accretion luminosity ($L_{\rm acc}$) using the empirical relation in~\citet{Fang2009A&A...504..461F}\footnote{There are also other empirical relations relating line luminosity to accretion luminosity \citep[e.g.,][]{Alcala2017A&A...600A..20A,Alcala2014A&A...561A...2A}. We also test the results using these relations, and using a different relation does not impact our results.}. Finally, the mass accretion rate is determined using Equation~\ref{eq:Lacc_to_Macc}, 
\begin{equation}\label{eq:Lacc_to_Macc}
\dot{M}_{\rm acc}=\dfrac{L_{\rm acc}R_{\star}}{GM_{\star}(1-R_{\star}/R_{\rm in})},
\end{equation}
where $R_{\rm in}$ denotes the truncation radius of the disk, and is assumed to be $5R_{\star}$~\citep{Gullbring1998ApJ...492..323G}, and $G$ is the gravitational constant. The stellar radius ($R_{\star}$) is obtained during the SED fitting, and the stellar mass ($M_{\star}$) is derived from the H-R diagram (see Section~\ref{sec:sedfitting}).

The mass accretion rates are compared to the stellar properties (stellar masses and ages) in Figure~\ref{fig:Macc_vs_star}. As shown in the left panel, the young population have accretion rates consistent with predicted from the relation in~\citet{Hartmann2016ARA&A..54..135H}, though with large scatter. But for the old population, the accretion rates are systematically lower than predicted from the empirical relation. The evolution of mass accretion rates with stellar ages is displayed in the right panel. Both young and old populations follow the trend in \citet{Hartmann2016ARA&A..54..135H} corrected for stellar masses (dashed line in the plot).

%%%%%%%%%%%%%%%%%%%%%%%%%%%%%%%%%%%%%%%%%%%%%%%%%%%%%%%%%%%%%%%%%%%%%%%%%%%%%%%%%%%%%%%%%%%%

\subsection{Disk Category}\label{sec:disktype}

\begin{figure*}[!t]
    \centering
    \includegraphics[width=\textwidth]{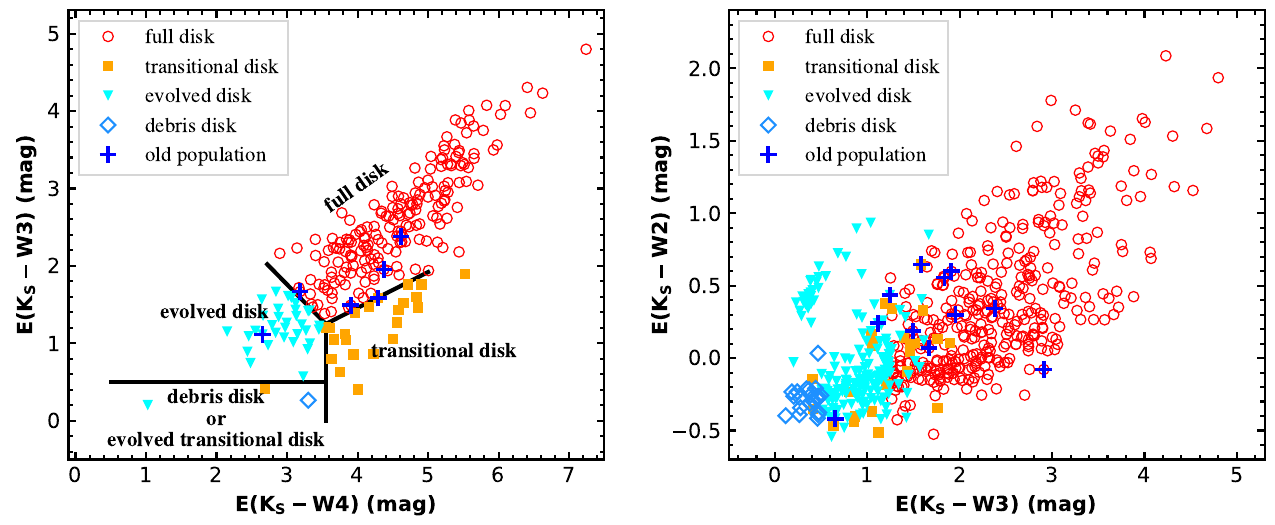}
    \caption{Left: $E(K_{S}-W3)$ vs. $E(K_{S}-W4)$ color excess plot for our targets. Full disks, transitional disks, evolved disks and debris disks are shown as red circles, orange squares, cyan triangles and blue diamonds, respectively. The newly discovered Peter Pan disks are highlighted with additional blue pluses. The solid lines display the boundaries used to classify our targets into different categories \citep{Esplin2014ApJ...784..126E,Esplin2018AJ....156...75E}. Right: similar as the left panel, but shows the  $E(K_{S}-W2)$ vs. $E(K_{S}-W3)$ color excess plot.}
    \label{fig:color_excess}
\end{figure*}

Infrared excess emission arising from circumstellar disks surrounding YSOs varies with the evolutionary stages. There are a variety of observational diagnostics to separate YSOs or circumstellar disks into different categories and evolutionary stages \citep[e.g.][]{Lada1984ApJ...287..610L,Greene1994ApJ...434..614G,Robitaille2006ApJS..167..256R,Robitaille2007ApJS..169..328R,Esplin2014ApJ...784..126E}. Many recent studies \citep[e.g.,][]{Fang2023ApJ...945..112F,Haerken2024ApJ...960...58H} adopted the scheme of \citet{Esplin2014ApJ...784..126E,Esplin2018AJ....156...75E}, using extinction corrected IR color excess, to classify YSOs into full, transitional, evolved, and debris or evolved transitional disks. In this work, we adopt the scheme of~\citet{Esplin2014ApJ...784..126E,Esplin2018AJ....156...75E}, using extinction corrected IR color excess, to classify the sources into full, transitional, evolved, and debris or evolved transitional disks. For sources have reliable photometry in $W3$ and $W4$-bands, they are classified using the $E(K_{S}-W3)$ vs. $E(K_{S}-W4)$ color excess plot (left panel in Figure~\ref{fig:color_excess}). For sources detected in $W3$ but not $W4$ band, we classify as full disks these have $E(K_{s}-W3)>1.25$, as evolved or transitional disks these have $0.5<E(K_{S}-W3)<1.25$, and as debris disks these have $E(K_{S}-W3)<0.5$ \citep{Luhman2022AJ....163...25L}. If a source lacks reliable photometry in both $W3$ and $W4$-bands but shows excess in $W1$ or $W2$-bands, we classify it as a full disk\citep{Luhman2012ApJ...758...31L,Luhman2022AJ....163...25L}. We note that several sources are classified as debris disks based on $E(K_{S}-W3)$ or $E(K_{S}-W4)$ but show weak excess in $W1$ or $W2$ bands, so they are reclassified as evolved disks. We also note one source (\textit{Gaia DR3 3417967688925113344}) is classified as a debris disk, showing significant excess in $W3$ and $W4$-bands, but have strong and broad H$\alpha$ emission line ($\rm EW_{H\alpha}=65\;\AA$), and display stochastic lightcurve\footnote{Stochastic variability is generally related to accretion activity \citep{Wang2023RAA....23g5015W} and \citet{Stauffer2016AJ....151...60S} attributed the variability to continuously stochastic accretion events producing transient hot spots on the stellar surface.}, so it is reclassified as an accreting transitional disk. There are 631 full disks, 179 evolved disks, 23 transitional disks, and 22 debris disks in our working sample. Following the prescription in \citet{Luhman2012ApJ...758...31L,Luhman2020AJ....160...44L,Luhman2022AJ....163...25L}, full disks, evolved disks and transitional disks are considered as protoplanetary disks. Finally, we identify 833 protoplanetary disks in our working sample. It is not surprising that most objects in our working sample are protoplanetary disks, since our main goal is to identify protoplanetary disks (see Section~\ref{sec:select_disk}). We also display the $E(K_{S}-W2)$ vs. $E(K_{S}-W3)$ color excess plot in the right panel of Figure~\ref{fig:color_excess} to illustrate the sizes of the excesses in $W2$-band for different disk categories.

%%%%%%%%%%%%%%%%%%%%%%%%%%%%%%%%%%%%%%%%%%%%%%%%%%%%%%%%%%%%%%%%%%%%%%%%%%%%%%%%%%%%%%%%%%%%

\section{Discussion}\label{sec:discussion}

\subsection{Validity of the Selection Method}\label{sec:validity}

Peter Pan disks are generally discovered as anomalies in studying stellar aggregates or star clusters that are older than typical lifetimes of protoplanetary disks. The oldness of the disks are assessed through associating them to associations or clusters with well determined ages. In this work, we perform a relatively blind search for long-lived protoplanetary disks, and the ages are determined individually for each source.

In this section, we discuss the validity of the selection method adopted in this work. We use the 15 Peter Pan disks listed in \citet{Lee2020MNRAS.494...62L} as comparison sample. To assess their properties through H-R diagram, we collect the stellar parameters (effective temperature and bolometric luminosity) from various literature \citep{Rodriguez2014A&A...567A..20R,Murphy2015MNRAS.453.2220M,Murphy2018MNRAS.476.3290M,Lee2020MNRAS.494...62L}. For sources lacking determinations in literature, we collect their optical photometry from PS1 and supplement it with synthetic photometry from the Gaia low resolution spectra \citep{Gaia-Collaboration2023A&A...674A..33G}, and apply the same SED fitting procedure as for our sample to determine their stellar parameters. Ten of the sources are placed in the H-R diagram (green stars in the left panel of Figure~\ref{fig:HRD}). Among these Peter Pan disks, only the source \textit{LDS 5606 A} is observed by LAMOST. But this source is not identified as an aged disk due to its slightly younger isochronal age ($\sim$9\;Myr). \citet{Lee2020MNRAS.494...62L} listed this source as a member of the Columnba cluster with an age of 30-44\;Myr, but \citet{Schneider2019AJ....157..234S} reported it as a bona fide $\beta$ Pic member aged 12-25\;Myr. There are 4 sources appear significantly above the 10\;Myr isochrone. Of them, \textit{2MASS J05082729-2101444} and \textit{LDS 5606 B} are of uncertain membership, similar as the case for \textit{LDS 5606 A}. Sources \textit{2MASS J12392312-5702400} and \textit{2MASS J13373839-4736297} are associated with the 10-17\;Myr old Lower Centaurus Crux (LCC) cluster \citep{Rodriguez2011ApJ...727...62R,Schneider2012ApJ...757..163S,Murphy2015MNRAS.453.2220M}. From our analysis, some of these Peter Pan disks appear to be isochronally young. This is mainly due to the difficulty of determine isochronal ages for individual stars, since there are great discrepancy among different sets of evolutionary tracks. For example, when comparing with the evolutionary model of \citet{Baraffe2015A&A...577A..42B}, only one of the literature Peter Pan disks (\textit{WISE J080822.18-644357.3}) is determined to be older than 10\;Myr (cyan dashed line in the left panel of Figure~\ref{fig:HRD}). \citet{Murphy2018MNRAS.476.3290M} also noted large discrepancies of the isochronal ages of the source \textit{WISE J080822.18-644357.3} among different sets of isochrones. In summary, on the one hand, although we can not pick up all the Peter Pan disks efficiently during the H-R diagram assessment, we do identify some isochronally aged disks, and these disks may serve as supplements greatly improving the census of Peter Pan disks. On the other hand, there exist some isochronally young but dynamically old Peter Pan disks. Many reasons maybe responsible for this discrepancy, such as that there are significant mass dependence in age estimation, especially for low mass stars \citep{Herczeg2015ApJ...808...23H}. Considering that some of the newly discovered aged YSOs are associated with open clusters (see Section~\ref{sec:spatial}), we may discover additional aged, but isochronally young YSOs through analyzing members of these clusters in the future.

%%%%%%%%%%%%%%%%%%%%%%%%%%%%%%%%%%%%%%%%%%%%%%%%%%%%%%%%%%%%%%%%%%%%%%%%%%%%%%%%%%%%%%%%%%%%

\subsection{Comparison with Previously Discovered Peter Pan Disks}

Previous discovery of Peter Pan disks are mainly through studying individual sources in specific associations, and the source age is determined to be the age of the corresponding association. In this work, we perform a blind search for aged YSOs across the whole area surveyed by the LAMOST. The Peter Pan disks identified here are not associated with any nearby associations, instead they are associated with open clusters or are isolated (see the discussion in Section~\ref{sec:spatial}).

Previously discovered Peter Pan disks are mid-M types, while the aged YSOs identified in this work include early-M and late-K types as well. As already mentioned in Section~\ref{sec:acc}, the Peter Pan disks identified in this work accretes at lower levels compared to the young population with similar stellar masses. We determine stellar masses and mass accretion rates for the Peter Pan disks listed in \citep[their Table~4]{Lee2020MNRAS.494...62L} using the same method as for our sample. Similar as for the aged YSOs identified here, these Peter Pan disks display lowered level of mass accretion rates as well (green stars in the left panel of Figure~\ref{fig:Macc_vs_star}). We find a common characteristic of lowered level of mass accretion rates for long-lived protoplanetary disks.

While most of the Peter Pan disks tabulated in \citet{Lee2020MNRAS.494...62L} are $<$50\;Myr, we identify several objects located well below the 50\;Myr isochrone. Though we could not rule out the possibility of them being edge-on disks and more observations are required to assess their properties, this discovery would be significant challenge to present theories of disk evolution. Although these sources are significantly aged, the evolution of mass accretion rates with stellar ages follows the same trend as the young population.

%%%%%%%%%%%%%%%%%%%%%%%%%%%%%%%%%%%%%%%%%%%%%%%%%%%%%%%%%%%%%%%%%%%%%%%%%%%%%%%%%%%%%%%%%%%%

\subsection{Spatial Distribution}\label{sec:spatial}

\begin{figure*}[!t]
    \centering
    \includegraphics[width=\textwidth]{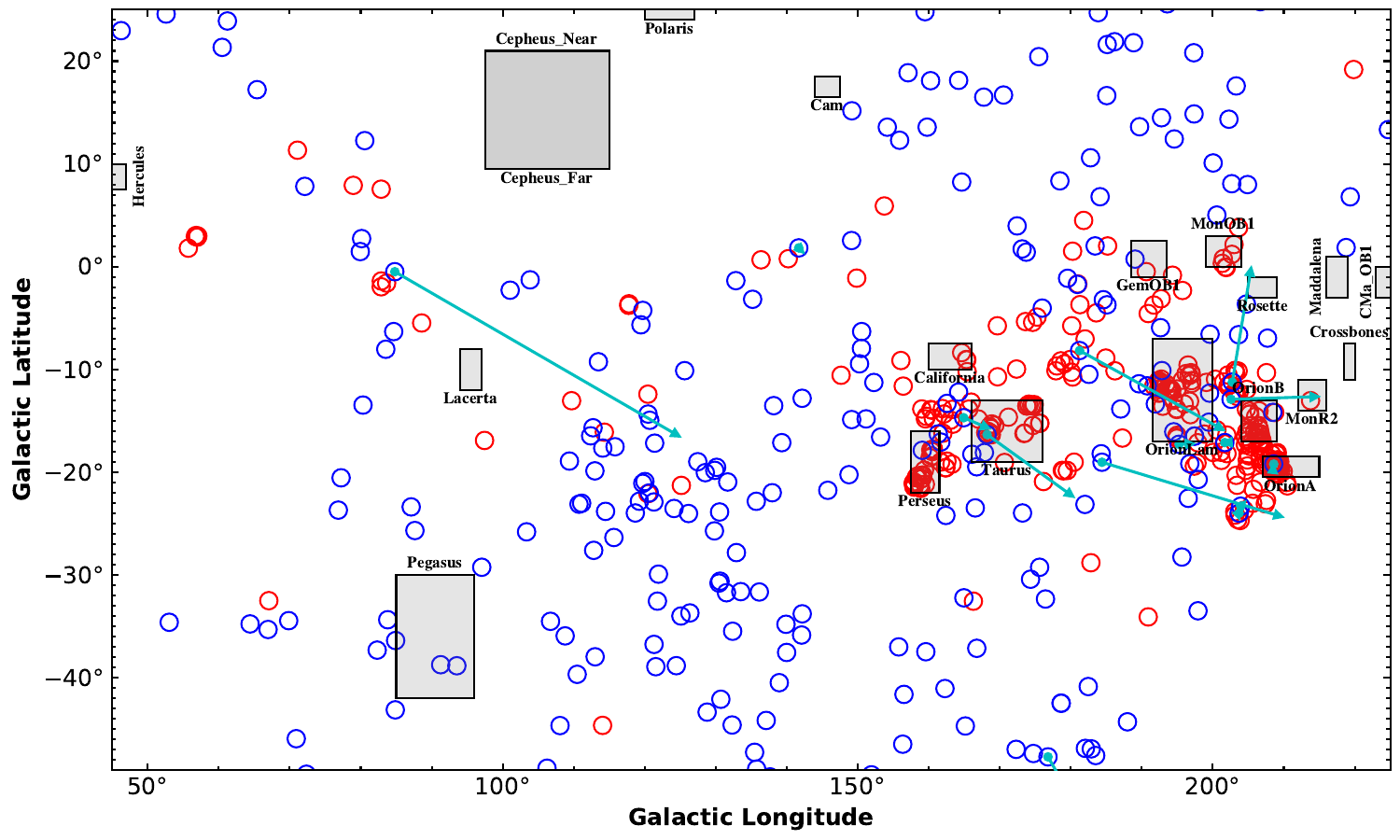}
    \caption{Spatial distribution of the disked objects studied in this work. Red and blue circles represent young and old populations respectively, and the Peter Pan disks identified in this work are marked with additional cyan dots with arrows denote their proper motions. The gray filled rectangles represent the sky coverage of nearby star forming regions~\citep{Zucker2019ApJ...879..125Z}.}
    \label{fig:stars_into_cca}
\end{figure*}

Previously discovered Perter Pan disks are generally associated with nearby associations, but most of the newly discovered ones are not. In Figure~\ref{fig:stars_into_cca}, we display the spatial distribution of the disked objects studied in this work. As demonstrated, most of the disked objects are projected toward or near star-forming regions \citep{Zucker2019ApJ...879..125Z}. While most of the young population are projected toward or near star-forming regions, nearly all the old population are not. Detailed comparison indicates that about half of the young population (151/329) also have distances consistent with the cloud, while only two of the old population do. We also compare the spatial distribution of our working sample with the cluster catalog of \citet{Hunt2023A_A...673A.114H}, and find that more than half of the young population (174/329) are clustered while only 2\% (12/526) of the old population show evidence of being members of open clusters.

Among the 14 new Peter Pan disks, only one has distance consistent with the Taurus region and only two are listed as members of the 32~Ori association aged $\sim$20\;Myr by \citet{Luhman2022AJ....164..151L}. The others are either associated with open clusters \citep{Hunt2023A_A...673A.114H} or isolated. One of the two association members is also listed as an open cluster member. Some of these isolated Peter Pan disks have very large proper motions, and they could be runaway stars from their parent clouds. Future observations of their radial velocity should enable trace-back analysis to determine their origins. There are also some stars in the young population appear isolated. Considering their young ages, they don't have enough time to move far away from their birth place, and they could be formed in situ, representing the type of isolated star formation. In fact, there are more and more observational evidence demonstrate that ``Not all stars form in clusters'' \citep{Ward2018MNRAS.475.5659W,Ward2020MNRAS.495..663W}.

In summary, while previously discovered Peter Pan disks are members of nearby associations, many of the newly discovered ones are associated with open clusters. Since more and more open clusters are continuing to be discovered with the Gaia mission, it is possible to discover more such cases in the future, through analyzing the cluster members carefully.

%%%%%%%%%%%%%%%%%%%%%%%%%%%%%%%%%%%%%%%%%%%%%%%%%%%%%%%%%%%%%%%%%%%%%%%%%%%%%%%%%%%%%%%%%%%%

\clearpage

\section{Summary}\label{sec:summary}

While both simulations and observations demonstrate rapid disk dissipation, several examples of prolonged accretion disks surrounding M stars have been observed. In this work, we carried out a systematic search for isochronally old M stars still surrounded by primordial disks in the LAMOST M star catalog. The main results are summarized as follows.
\begin{enumerate}
\item In this work, we studied 855 disked objects, including 526 isochronally old and 329 isochronally young stars, identified from the LAMOST M star catalog. The stellar parameters, and the variability and accretion properties are determined for these disked objects. Most of the young disked objects are variables (90\% of the population) and strong accretors (64\% of the population), but for the old population, these frequencies are fairly low (21\% and 3\% of the populations respectively).
\item Through analyzing the source properties, we identify 14 objects as Peter Pan disks, nearly doubled the category. All but 3 of these new Peter Pan disks are variables. We find systematically lowered levels of mass accretion rates for the Peter Pan disks compared to their young counterparts with similar masses, but the evolution of mass accretion rates with stellar ages follows similar trend as young accretors.
\item Unlike previously discovered Peter Pan disks that are associated with nearby associations, the newly discovered ones are members of open clusters or are isolated.
\end{enumerate}

%%%%%%%%%%%%%%%%%%%%%%%%%%%%%%%%%%%%%%%%%%%%%%%%%%%%%%%%%%%%%%%%%%%%%%%%%%%%%%%%%%%%%%%%%%%%

\begin{acknowledgments}
This study is supported by the National Natural Science Foundation of China under grant No. 12173013; the project of Hebei Provincial Department of Science and Technology under grant number 226Z7604G, and the Hebei NSF (No. A2021205006). Xiao-Long Wang acknowledge the support by the Science Foundation of Hebei Normal University (Grant No. L2024B56) and S\&T Program of Hebei (Grant No. 22567617H)). Miao-miao Zhang acknowledges the support by the National Natural Science Foundation of China (grants No. 12073079). YL acknowledges financial supports by the National Natural Science Foundation of China (Grant number 11973090), and by the International Partnership Program of Chinese Academy of Sciences (Grant number 019GJHZ2023016FN). This work has made use of data from the European Space Agency (ESA) mission {\it Gaia} (\url{https://www.cosmos.esa.int/gaia}), processed by the Gaia Data Processing and Analysis Consortium (DPAC, \url{https://www.cosmos.esa.int/web/gaia/dpac/consortium}). Funding for the DPAC has been provided by national institutions, in particular the institutions participating in the Gaia Multilateral Agreement. Guoshoujing Telescope (the Large Sky Area Multi-Object Fiber Spectroscopic Telescope LAMOST) is a National Major Scientific Project built by the Chinese Academy of Sciences. Funding for the project has been provided by the National Development and Reform Commission. LAMOST is operated and managed by the National Astronomical Observatories, Chinese Academy of Sciences.
\end{acknowledgments}

\clearpage

\appendix

% \section{}
% \setcounter{figure}{0}

\section{Examples of SEDs and WISE Images Demonstrating the Visual Inspection}\label{app:wise_image}

\begin{figure*}[!t]
    \centering
    \includegraphics[width=\textwidth]{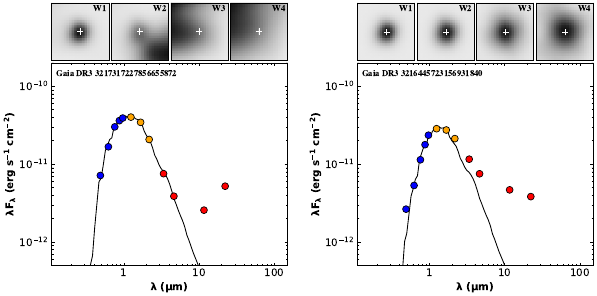}
    \caption{Left: SED and WISE images of a selected candidate that is rejected by our visual inspection. The top panels show the WISE images of this source and the target location is marked as white plus in each panel. The bottom panel shows the SED of the candidate. Blue, orange and red circles are optical, NIR and WISE photometry, respectively. The solid line represents the best-fitted stellar photosphere (see Section~\ref{sec:sedfitting}). Right: similar as the left panels, but for a source passing our visual inspection.}
    \label{fig:SED_and_WISEimage}
\end{figure*}

\citet{Koenig2014ApJ...791..131K} noted that non-negligible fraction of the sources listed in the AllWISE source catalog are fake source contamination, resulting in unreliable photometry. The contamination rates are relatively low in $W1$ and $W2$ bands, but can exceed 95\% in $W4$ band. Although we have removed photometry with magnitude errors $\ge0.2\;\rm mag$, some of the identified disked candidates may still be fake source contamination, especially when there are only excess emission in $W3$ or $W4$ bands. For this reason, WISE images and SEDs are visually inspected one-by-one to remove unreliable photometry due to fake source detection.

In this appendix, we display two examples (Figure~\ref{fig:SED_and_WISEimage}) to demonstrate the necessity of inspecting their WISE images for reliably identifying disked objects. If only the SEDs are considered, both sources display significant excess emission in $W3$ and $W4$ bands. But as already mentioned in Section~\ref{sec:select}, for the source displayed in the left panel, there are no point sources at the target's location. Thus the $W3$ and $W4$ photometry for this source are due to fake source detection, resulting in erroneous identification of this source as a disked object. As comparisons, the source displayed in the right panel is well shaped in all four WISE bands, and its identification as a disked object is trustworthy.

%%%%%%%%%%%%%%%%%%%%%%%%%%%%%%%%%%%%%%%%%%%%%%%%%%%%%%%%%%%%%%%%%%%%%%%%%%%%%%%%%%%%%%%%%%%%

\section{Catalog of the Whole Working Sample of Disked Objects}\label{app:AllDiskedObjects}

In this appendix, we provide the full information for the whole working sample of disked objects, including the 14 Peter Pan disks in Table~\ref{tab:AllDiskedObjects}.

\begin{table*}[!t]
\centering%
\renewcommand{\arraystretch}{1.1}\setlength{\tabcolsep}{1.5ex}
\caption{Column overview of the catalog containing all the disked objects studied in this work.}\label{tab:AllDiskedObjects}
\begin{tabular}{llp{0.68\textwidth}}\hline\hline
Column Name             & Unit               & Description\\\hline
GaiaDR3                 & ---                & Gaia DR3 source id\\
RA                      & deg                & Right ascension (J2016)\\
DEC                     & deg                & Declination (J2016)\\
parallax                & mas                & Parallax\\
parallax\_error         & mas                & Parallax error\\
pmra                    & $\rm mas\;yr^{-1}$ & proper motion in RA direction\\
pmra\_error             & $\rm mas\;yr^{-1}$ & Error in pmra\\
pmdec                   & $\rm mas\;yr^{-1}$ & proper motion in DEC direction\\
pmdec\_error            & $\rm mas\;yr^{-1}$ & Error in pmdec\\
SPT                     & ---                & Spectral type\\
eSPT                    & ---                & Error in the spectral type\\
$T_{\rm eff, 0}$        & K                  & Effective temperature obtained by fitting stellar photosphere\\
$\log L_{\rm bol, 0}$   & $L_{\odot}$        & Bolometric luminosity obtained by fitting stellar photosphere\\
$T_{\rm eff, 1}$        & K                  & Effective temperature obtained by fitting disked model\\
$\log L_{\rm bol, 1}$   & $L_{\odot}$        & Bolometric luminosity obtained by fitting disked model\\
$M_{\star}^{(a)}$       & $M_{\odot}$        & Stellar mass\\
Age$^{(a)}$             & Myr                & Stellar age\\
HaEM                    & ---                & $=$1 for H$\alpha$ emitters\\
$\rm EW_{H\alpha}$      & \AA                & Equivalent width of H$\alpha$ emission line\\
$\log\dot{M}_{\rm acc}$ & $M_{\odot}\;\mathrm{yr}^{-1}$ & Mass accretion rate\\
TTS                     & ---                & $=$\texttt{CTTS} for accretors, and $=$\texttt{WTTS} for non-accreting stars\\
variable                & ---                & $=$Y for variables, and $=$N for non-variables\\
DiskType                & ---                & Disk classifications, with \texttt{FULL} for full disks, \texttt{EVOLVED} for evolved disks, \texttt{TD} for transitional disks, and \texttt{DB} for debris disks\\
Cluster                 & ---                & The open cluster ID from \citet{Hunt2023A_A...673A.114H}\\
ClusterType             & ---                & $=$2 for sources projected toward the cluster, $=3$ for sources with distances consistent with the clusters as well, $=5$ for sources with proper motions also consistent with the clusters, $=6$ for sources listed as members by \citet{Hunt2023A_A...673A.114H}\\
Cloud                   & ---                & Molecular cloud name\\
CloudType               & ---                & $=2$ for sources projected toward the molecular clouds, $=$3 for sources with distances consistent with the molecular cloud as well\\
population              & ---                & $=$\texttt{old} for old population, and $=$\texttt{young} for young population\\
\hline
\multicolumn{3}{p{0.95\textwidth}}{$^{(a)}$ Stellar masses and ages are obtained from the H-R diagram constructed using $T_{\rm eff}$ and $L_{\rm bol}$ derived from fitting disked models.}\\
\multicolumn{3}{l}{(This table is available in its entirety in fits format.)}
\end{tabular}
\end{table*}

%%%%%%%%%%%%%%%%%%%%%%%%%%%%%%%%%%%%%%%%%%%%%%%%%%%%%%%%%%%%%%%%%%%%%%%%%%%%%%%%%%%%%%%%%%%%

\section{Examples Demonstrating the Spectral Typing}\label{app:spec_example}

\begin{figure*}[!t]
    \centering
    \includegraphics[width=\textwidth]{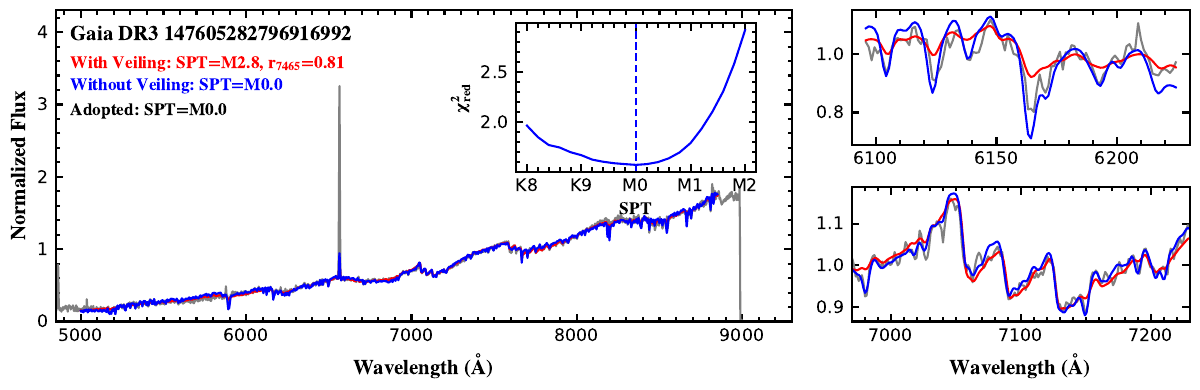}
    \caption{Illustration of our spectral typing for \textit{Gaia DR3 147605282796916992}, where the nonveiled model is the preferred fit to the observed spectrum. Left: best-fitted templates with (red) and without (blue) veiling overplotted on the observed spectrum (gray). The inset shows the distribution of the reduced $\chi^{2}$ derived from the fitting with the spectral types of the templates. Right: zoomed-in comparison of the target spectra and the best-fitted templates with (red) and without (blue) veiling in the wavelength range 6100$-$6220\;$\rm\AA$ and 7000$-$7220\;$\rm\AA$.}
    \label{fig:spec_example1}
\end{figure*}

\begin{figure*}[!t]
    \centering
    \includegraphics[width=\textwidth]{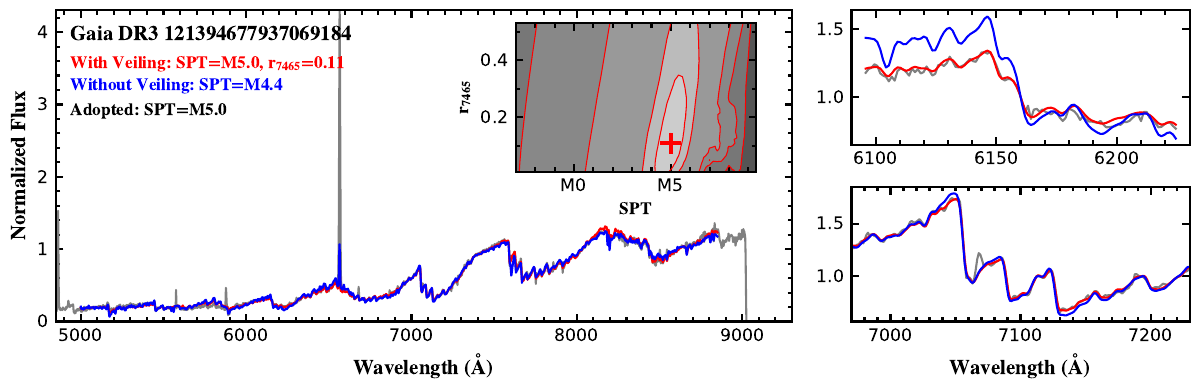}
    \caption{Same as Figure~\ref{fig:spec_example1}, but for the source \textit{Gaia DR3 121394677937069184}, which is better fitted with veiled template (red). The inset in the left panel displays the reduced $\chi^{2}$ as a function of different combinations of SPT and $r_{7465}$.}
    \label{fig:spec_example2}
\end{figure*}

In this appendix, we display two examples to demonstrate the spectral typing process. Figure~\ref{fig:spec_example1} is for a source that is better fitted with nonveiled template, and Figure~\ref{fig:spec_example2}  displays a source that is better fitted with veiled template.

%%%%%%%%%%%%%%%%%%%%%%%%%%%%%%%%%%%%%%%%%%%%%%%%%%%%%%%%%%%%%%%%%%%%%%%%%%%%%%%%%%%%%%%%%%%%
%%%%%%%%%%%%%%%%%%%%%%%%%%%%%%%%%%%%%%%%%%%%%%%%%%%%%%%%%%%%%%%%%%%%%%%%%%%%%%%%%%%%%%%%%%%%
%%%%%%%%%%%%%%%%%%%%%%%%%%%%%%%%%%%%%%%%%%%%%%%%%%%%%%%%%%%%%%%%%%%%%%%%%%%%%%%%%%%%%%%%%%%%

\bibliography{References}{}
\bibliographystyle{aasjournal}

\end{document}